\documentclass{iaus}
\usepackage{graphicx}
\usepackage{natbib}

\title[Stars in the age of micro-arc-second astrometry] 
{Stars in the age of micro-arc-second astrometry}

\author[Yveline Lebreton]   
{Yveline Lebreton  
  }

\affiliation{GEPI, UMR 8111, Observatoire de Paris-Meudon, 92195 Meudon, France\break email: Yveline.Lebreton obspm.fr}

\pubyear{2008}
\volume{248}  
\pagerange{119-127}
\date{?? and in revised form ??}
\jname{Proceedings Title IAU Symposium}
\editors{A.C. Editor, B.D. Editor \& C.E. Editor, eds.}
\begin{document}

\maketitle

\begin{abstract}

The understanding and modeling of the structure and evolution of stars is based on statistical physics as well as on hydrodynamics. Today, a precise identification and proper description of the physical processes at work in stellar interiors are still lacking (one key point being that of transport processes) while the comparison of real stars to model predictions, which implies conversions from the theoretical space to the observational one, suffers from uncertainties in model atmospheres. That results in uncertainties on the prediction of stellar properties needed for galactic studies or cosmology (as stellar ages and masses). In the next decade, progress is expected from the theoretical, experimental and observational sides. I illustrate some of the problems we are faced with when modeling stars and the possible tracks towards their solutions. I discuss how future observational ground-based or spatial programs (in particular those dedicated to micro-arc-second astrometry, asteroseismology and interferometry) will provide precise determinations of the stellar parameters and contribute to a better knowledge of stellar interiors and atmospheres in a wide range of stellar masses, chemical compositions and evolution stages.

\end{abstract}

\section{Stellar internal structure and evolution studies: goals and tools}

Major goals of stellar structure and evolution studies are (i) to characterize and describe the physics of matter in the extreme conditions encountered in stars and (ii) to determine stellar properties (like age and mass) that trace the history and evolution of galaxies and constrain cosmological models. To achieve these goals, we rely on numerical stellar models based on input physics that integrate the results of recent theoretical studies, numerical simulations and laboratory experiments. The models inputs and outputs are chosen and/or validated by comparison with accurate astronomical observations.

Numerical 2 and 3{\small D} hydrodynamical simulations of limited regions of stellar interiors and atmospheres are now under reach of computers. They provide valuable constraints and data for current standard (1{\small D}) stellar models: abundances, convection, rotationally induced instabilities and mixing, magnetic fields, etc. \citep[see e.g.][for reviews]{2005ARA&A..43..481A,2007arXiv0708.1499T,2007IAUS..239..517Z}.
In parallel, the physics of stellar plasmas is studied in the laboratory with (i) fluid experiments \citep[study of turbulence in rotating, magnetic fluids, etc., see e.g.][]{1999A&A...347..734R}, (ii) particle accelerators (nuclear reaction cross sections, etc.) and, (iii) the so-called high energy-density facilities (based on high power lasers or z-pinches) which aim at exploring the high temperature and high density regimes found in stars, brown dwarfs and giant planets to get  information on the equation of state ({\small EOS}), opacities or thermonuclear reactions \citep[see][for a review]{2006RvMP...78..755R}.

Modern ground-based and spatial telescopes equipped with high quality instrumentation are in use or under development ({\small VLT-VLTI, JWST,} etc.). They provide very accurate data which, after treatment, give access to stellar global parameters (luminosity, radius, mass, effective temperature $T_{\rm eff}$, gravity $\log g$, abundances, etc.).
On the other hand, seismic data (such as oscillation frequencies or amplitudes) are being obtained in velocity from the ground \citep[see, e.g.][]{2007CoAst.150..106B} and in photometry by the space missions {\small MOST} \citep{2003PASP..115.1023W} and {\small CoRoT} \citep{2006corm.book...39M}. In the next decade, valuable observational data are expected. For instance {\small GAIA} \citep{ESA2000,2001A&A...369..339P}, to be launched in 2011, will make astrometric measurements, at the micro-arc second level together with photometric and spectroscopic observations of a huge number of stars covering the whole range of stellar masses, compositions and evolution stages while the Kepler mission, to be launched in 2009, will provide the opportunity to make asteroseismic observations on a wide range of stars \citep{2007CoAst.150..350C}.

In the following, I discuss the different aspects of stellar modeling, the problems encountered and the perspectives.

\section{Stellar models: input parameters and observational constraints}
\label{models}

\subsection{Input physics for stellar models}

Stellar model calculation requires a good description of the physical processes at work. Microscopic physics (opacities, {\small EOS}, nuclear reaction rates, atomic diffusion) are now rather well described which improves the agreement between models and observations. However difficulties still arise in the modeling of e.g. (i) cold, dense stars (molecular opacities, non-ideal effects in the {\small EOS}), (ii) advanced stages of evolution (nuclear reaction rates) or (iii) stars to be modeled very accurately like the Sun (see Sect.~\ref{solar-model}).
On the other hand, despite important recent progress, macroscopic processes (convection, transport of chemicals and angular momentum related to differential rotation and the role of magnetic fields and internal gravity waves) are not fully understood  \citep[see e.g.][]{2007arXiv0708.1499T}.

\subsection{Inputs from observations and from model atmospheres}

Stellar models calculation involves inputs and constraints derived from observation. Observational data (e.g. magnitudes, colors, spectra, light and velocity curves, astrometric data) must be treated to get stellar parameters (luminosity, mass, radius, abundances, etc.). Model atmospheres are a crucial step in the analysis, for instance to predict fluxes in different bands, synthetic spectra or limb-darkening coefficients. They also provide outer boundary conditions for the interior model as well as bolometric corrections and color-temperature conversions (to convert the outputs of interior models into quantities comparable with observations). 
Recently, model atmospheres have been improved by the bringing-in of better atomic and molecular data and of modern computers and algorithms. The outputs of 2-3{\small D MHD} simulations, including {\small NLTE} effects, begin to be used to derive abundances, $T_{\rm eff}$ or $\log g$ from high quality spectra \citep{2005ARA&A..43..481A,2007IAUS..239..517Z}.

However, although the internal errors in the determination of stellar parameters are becoming quite small for dwarfs, subgiants and giants of spectral types A to K \citep[see Table~1 in][]{2005ESASP.576..493L}, large systematic errors remain for cool and hot stars. For instance, the systematic errors on the metallicity of metal poor dwarfs and cool giants amount to 0.2-0.3 dex, that is ten times more than the typical internal errors, while differences in $T_{\rm eff}$-scales can reach 200 to 400~K \citep{gustaf04,2006ASPC..352..105A}.

Oscillations have been detected in many stars: solar-like, $\delta$ Scuti, $\beta$ Cephei, $\gamma$ Dor, Cepheids, RR Lyrae, SPB, WD, etc.  They result from the propagation of acoustic pressure waves or of gravity waves, depending on the mass, evolution stage, chemical composition and excitation mechanism. Valuable constraints can be drawn from the frequencies or their combination \citep[see, e.g.][and references therein]{1988IAUS..123..295C,2003A&A...411..215R}.

\section{Modeling calibrators}

The rather few stars for which we have strong or numerous observational constraints are used as calibrators, i.e. they serve to validate the models and learn on the physics. What is learned from calibrators can then be applied to stars with incomplete or less accurate observations. This implies to extrapolate to compositions, masses or evolution stages not covered by calibrators. Quantities of astrophysical interest such as age, helium content or distance scale can then be derived for very large numbers of stars.

\subsection{Solar modeling: AGS05 mixture and seismology}
\label{solar-model}

The solar photospheric abundances have been re-determined recently on the basis of 3{\small D} radiative-hydrodynamical model atmospheres including better atomic data and {\small NLTE} effects \citep[see][]{2005ARA&A..43..481A}. The new mixture, referred to as  {\small AGS05}, gives a global metallicity and abundances of {C, N, O} smaller by $30$-$40\%$ than those given by the {\small  GN93} mixture \citep{GN93} derived from 1{\small D} hydrostatic models. As a result, the interior opacity is reduced and the solar model no more satisfies helioseismic constraints (convection zone depth, sound speed profile, etc.), the problem being especially acute in the region between  $\sim$0.4 and $\sim$0.7$R_\odot$ (upper part of the radiative zone and shear region below the convective zone, i.e. tachocline). Several authors have shown that the present uncertainties of the input physics of the solar model, in particular the opacities or the atomic diffusion velocities, can hardly explain the differences \citep[see e.g.][]{2006CoAst.147...80M}. On the other hand, it has been suggested that the neon abundances could be in error by a factor of at least 2, but the problem is still open \citep[][]{2007SSRv..tmp..105G}.

\subsection{Binary system modeling: the RS Cha and $\alpha$ Centauri binary systems}

The modeling of a binary system consists in reproducing the observed constraints under the assumption that the two stars have same age and initial composition. This may allow to infer the values of the model unknowns: age, initial helium, and physical parameters as the mixing-length parameter for convection or overshooting parameter.

The binary system {\small RS} Cha is an interesting eclipsing, {\small SB2} system whose components are A-type oscillating stars in a {\small PMS} evolution stage corresponding to the onset of the earliest reactions of the {\small CNO} cycle. The modeling of the system by \citet{2007A&A...465..241A, 2007A&A...473..181A} at the light of new accurate observations (masses, radii, metallicity) has shown that, to get an agreement between models and observations, carbon and nitrogen must be depleted with respect to their values in the {\small GN93} mixture. The {\small AGS05} mixture fulfills this condition but to assess this result and the values of the system age and helium abundance derived from the calibration, it is now necessary to further improve the observational data. In particular, the {\small $\rm [Fe/H]$} value should be redetermined using 3{\small D} model atmospheres and it would be valuable to get individual abundances of major elements and better seismic data, the present ones being too coarse to provide useful constraints.

The binary system $\alpha$ Centauri is the closest and best-known one. The observed global parameters are accurate (in particular the accuracy on the interferometric radii is of $1\%$) and seismic observations of both stars have allowed to measure the frequencies of  several low degree p-modes with an accuracy $\sigma_\nu\simeq\rm 0.3$-$2.0~\mu{Hz}$. Several authors have performed a calibration of the system \citep[see, e.g.][and references therein]{2005A&A...441..615M}.
They conclude that it is difficult to find a set of parameters that satisfies simultaneously the global and seismic constraints. As a result, the constraints on the physics of the models remain loose and age and initial helium abundance are still poorly determined. To progress, it would be interesting to better assess the radii and masses by confirming the parallax of the system (different values were obtained from the analysis of Hipparcos data) and to further improve and enlarge the seismic data that are still scarce and coarse.

\subsection{Modeling stars in open clusters}

Members of stellar clusters can also serve as calibrators. They can be studied under the assumption that they all have the same age and initial composition but different mass. The initial helium abundance and age of a cluster can be derived from the comparison of a model isochrone with observations of cluster stars in a color-magnitude diagram \citep[see Fig. 3 in][]{2005ESASP.576..493L}. The surface He abundance $Y$ can be derived from the position of the lower main sequence ({\small MS}) and the age from the {\small MS} turn-off. Such studies require accurate observations (parallax, magnitude, $T_{\rm eff}$, $\rm [Fe/H]$). The metallicity uncertainty affects the $Y$ estimate because of the helium-metallicity degeneracy in the H--R diagram. Also, uncertainties on $Y$ and on age come from the bad knowledge of input physics, as envelope or core convection, rotational mixing or atomic diffusion \citep{2001A&A...374..540L}. 

The observation of binary stars in a cluster can provide additional constraints. For instance, the position of the low-mass, non-evolved stars in the mass-luminosity plane is related to their helium abundance while physics can be constrained if several binaries spanning a large mass range can be observed. Again the accuracy on the parallax and metallicity are crucial. A study of the Hyades (the only cluster where individual distances have been obtained by Hipparcos and masses measured for a few stars) has shown the limitation of the method due to the uncertainties on the metallicity and input physics and to the small number of stars with accurate mass determination \citep{2001A&A...374..540L}.

Investigations by e.g. \citet[][]{2001A&A...377..192M, 2004MNRAS.350..277B,2005A&A...430..571P} have shown that the seismic analysis of different kinds of oscillators should help to probe their inner properties as the outer convection zone depth and helium content or the convective core boundary and to estimate their mass and age. For instance, in solar-type stars, the higher the helium abundance in the convective envelope, the deeper the depression in the adiabatic index $\Gamma_1$ in the region of second helium ionisation. As shown by \citet{2004MNRAS.350..277B}, the helium abundance in the envelope of low-mass stars could be derived using the signature of this depression in the p-mode frequencies. This would require that low degree p-modes are observed with a frequency accuracy of $0.01\%$ and that the mass or radius of the star is known independently.

\section{Deriving astrophysical parameters: the example of stellar ages}

In our Galaxy, the ages of A and F stars are crucial inputs for studying the disc while those of old metal poor stars and  globular clusters provide valuable constraints for cosmology. The uncertainty on age depends on many factors (precision of the observed position in the H--R diagram, abundances, knowledge of  model input physics as convection, rotational mixing, atomic diffusion). The case of globular clusters is discussed by Chaboyer (these proceedings). I focus here on the ages of A-F stars.

\subsection{Ages of A and F stars and the size of their mixed cores}

A-F stars have convective cores on the {\small MS} and may be fast rotators in the $\delta$ Scuti instability strip. Therefore to model these stars, we are faced with difficulties in describing central extra-mixing by core convection overshoot, rotationally induced mixing in the radiative zone and we have to estimate the effects of rotation on photometric data. These processes modify either the stellar models or the position of the observed star in the H--R diagram and in turn affect the age determination.

The efficiency of mixing in the stellar core determines the quantity of fuel available to the star with crucial return on its lifetime. Overshooting of the convective cores produces an extra-mixing. In model calculation, this extra-mixing is usually crudely parametrised with a coefficient $\alpha_{\rm ov}$, the value of which probably depends on mass, composition and evolution stage as shown by empirical calibrations based on binaries and {\small MS} width observations \citep[][]{2000MNRAS.318L..55R,2001ApJ...556..230Y,2002A&A...392..169C}. Rotationally induced mixing can also bring extra fuel to the stellar engine. Recently, progress has been made in the current modeling of  the transport of angular momentum and chemicals which results from differential rotation \citep[see the review by][]{2007EAS....26...65M}.

\citet{2002ASPC..259..306G} have calculated models of a typical A-star including either overshooting or rotational mixing and have found that these distinct processes cannot be discriminated in the H--R diagram where they have similar signatures. \citeauthor{2002ASPC..259..306G} also showed that if the A-star pulsates as a $\delta$-Scuti star then the signature of the mixing process could be seen in the oscillations frequencies, provided enough modes are observed and identified. Today we estimate that an uncertainty on age ranging from $13$ to $24\%$ results from the poor knowledge of the inner mixing processes \citep{1995IAUS..166..135L}. We expect that in the near future, the improvement on the observed H--R diagram positions (in particular on the luminosity with micro-arc-second astrometry) and the availability of precise seismic data will allow to reduce this uncertainty to $3$-$5\%$ \citep{2005ESASP.576..493L}.

\section{Perspectives in the context of micro-arc-second astrometry}

Today very few calibrators are available to probe the physics of stellar interiors. After Hipparcos, $\sim$200 stars have distances accurate to better than $1\%$. Also, the sample of stars with masses and radii measured with an accuracy better than $1\%$ remains small. Concerning open clusters, Hipparcos provided precise individual distances only in the Hyades while binaries have been analysed in the Hyades and the Pleiades only. There is no cluster star where solar-like oscillations have been detected. Concerning A-F stars, Hipparcos determined the distances of $10^3$ A-F stars with an accuracy better than $\sim$10\% while seismic data have been obtained for quite many stars, but there  are often coarse.

In the next decade, we expect that observations of stars will increase both in numbers and in quality. Missions dedicated to global astrometry like {\small GAIA} \citep{ESA2000,2001A&A...369..339P} and {\small SIM} \citep{2007arXiv0708.3953U} will reach a parallax accuracy better than 10 micro-arcseconds. In parallel, the measurements of stellar parameters (magnitudes, temperatures, abundances, masses, radii etc.) are expected to be much improved thanks to high resolution spectroscopy, interferometry, etc. High-quality seismic data are expected from several missions: {\small CoRoT} will reach an accuracy of $0.1\mu\rm Hz$ on frequencies for about 50 targets (mainly solar-like oscillators, $\beta$ Cephei and $\delta$ Scuti) while  Kepler will  enlarge the seismic sample to thousands of stars with a frequency accuracy of $0.1$-$0.3\mu\rm Hz$. 

\subsection{Expected returns from {\small GAIA}} 

{\small GAIA} astrometry will be complemented by photometry and spectroscopy allowing most masses and evolution stages to be unprecedentedly documented. The number of calibrators of stellar physics will be drastically increased and homogeneous global parameters will be provided, e.g. magnitudes, masses and abundances. The parallax of $7~10^5 (21~10^6)$ stars will be measured with an accuracy of at least $0.1\% (1\%$) and the mass of stars in $17~000$ binary systems will be obtained with an accuracy better than $1\%$. About $120$ open clusters (up to 1~kpc) will be brought to a level of precision better than the one now reached in the Hyades and  $\sim$10 binaries per cluster will be observed. Parallax measurements, accurate to better than $0.5\%$, will be provided for $5~10^5$ A stars and $3~10^6$ F stars. Furthermore, while -after Hipparcos- direct distances are available with an accuracy better than 12$\%$ for only 11 subdwarfs and 2 subgiants, {\small GAIA} will provide (i) precise direct distances for very large samples of subdwarfs and for all subgiants up to 3kpc and (ii) individual distances with an accuracy better than 10$\%$ for stars in $~20$ globular clusters.
We will therefore work on very precise H--R diagrams of very large stellar samples with complementary data (as mass, radius, detailed abundances and seismic data) for various subsamples of stars. The interpretation of these data in the light of future improvements on theoretical, numerical and experimental physics will certainly bring further insight in the understanding of the stellar interiors and evolution.



\end{document}